\documentclass[aps, pra, reprint, superscriptaddress,longbibliography]{revtex4-2}
\usepackage[colorlinks,bookmarks=false,citecolor=blue,linkcolor=red,urlcolor=blue]{hyperref}
\usepackage{verbatim,float}
\usepackage{color}
\usepackage{amsmath,amssymb,bm,braket,float,mathtools,subfigure}
\usepackage{graphicx,xfrac,appendix}
\usepackage[english]{babel}
\usepackage{amsmath}
\usepackage[T1]{fontenc}
\usepackage[utf8]{inputenc}
\usepackage{graphicx}
\makeatletter
\usepackage{epstopdf,float}
\usepackage{epstopdf}
\makeatother

\begin{document}
\title{Unbound States and Mixed Bound--Unbound Phases in Near-Infinitely Deep Potentials}

\author{Shujie Cheng}
%\thanks{chengsj@zjnu.edu.cn}
\affiliation{Xingzhi College, Zhejiang Normal University, Lanxi 321100, China}
\affiliation{Department of Physics, Zhejiang Normal University, Jinhua 321004, China}

\author{Tong Liu}
\thanks{t6tong@njupt.edu.cn}
\affiliation{Department of Applied Physics,School of Science,Nanjing University of Posts and Telecommunications,Nanjing 210003, China}

\author{Gao Xianlong}
\thanks{gaoxl@zjnu.edu.cn}
\affiliation{Department of Physics, Zhejiang Normal University, Jinhua 321004, China}

\begin{abstract}
We investigate the robustness of unbound states in one-dimensional quasiperiodic models with near-infinitely deep potentials. By constructing a deeper extension of the Liu-Xia model and combining inverse participation ratio (IPR) calculations with Lyapunov-exponent analysis based on Avila's global theory, we show that increasing the potential depth does not eliminate unbound states. Instead, it shifts and narrows their energy window to $-2t-V<E<2t-V$. We further extend the analysis to non-Hermitian quasiperiodic potentials with gain and loss. In these systems, unbound states survive within analytically determined real-energy intervals, but they no longer occupy the whole interval uniformly; rather, they coexist with bound states and form a mixed bound-unbound phase. The corresponding boundaries between the mixed region and the pure bound-state regions are obtained exactly from the Lyapunov exponent. These results demonstrate that unbound states in extreme quasiperiodic potentials are controlled not only by the potential depth but also by the spectral and localization structures induced by non-Hermiticity.
\end{abstract}
\maketitle

\section{Introduction}
In the framework of quantum mechanics, the eigenstate properties of particles in potential wells
have always been a core research topic, and the classification of bound and unbound states is the
basis for understanding quantum transport and metal-insulator transitions \cite{Shankar1994}.
For a one-dimensional infinitely deep potential well, the textbook conclusion holds that the particle
can only exist in discrete bound states, and the unbound scattering states can only be realized in
a finite-depth or bounded potential field \cite{Stillinger1975,Marinica2008}. The traditional
definition divides the quantum states by the particle's eigenvalue relative to the potential barrier
height: when the eigenvalue is lower than the potential barrier, the particle is confined in the
potential well to form a bound state; when the eigenvalue is higher than the potential barrier,
the particle escapes the potential well to form an unbound state \cite{Capasso1992}, which is accompanied by the
correspondence between discrete spectrum and bound state, continuous spectrum and unbound state \cite{Molina2012,Gorkunov2020}.

However, the discovery of bound states in the continuous spectrum by von Neumann and Wigner
in 1929 broke the inherent cognition of the correspondence between quantum states and energy
spectra \cite{vonNeumann1929}, and a series of follow-up studies \cite{Weber1995,Weber1998} have found similar exotic
quantum states in various potential-field structures, which indicates the richness and complexity
of quantum state characteristics in non-traditional potential systems \cite{Koshelev2018,Kang2021}.
Until recently, Liu and Xia proposed a one-dimensional discrete quasiperiodic near-infinitely
deep potential model $V(n)=V\sec^2(2\pi\alpha n)$ ($V$ is the potential strength), which for the first time proved that unbound
states can exist in the potential field that tends to near-infinite depth \cite{Liu2023}. This
model is different from the traditional infinite deep potential well in finite space: its potential
strength approaches infinity only when the lattice site $n\to\infty$, and the
critical energy $E=2t$ ($t$ is the unit of energy)
is the mobility edge that separates bound and unbound states. When $E<2t$, the quantum state is in an
unbound state that spreads over the entire system space; when $E>2t$,
the quantum state forms a localized bound state \cite{Liu2023,Dwiputra2022}. This research 
breaks with the classical understanding of infinite deep potential wells and opens up a new research
direction for the study of unbound states in extreme potential fields.

On the one hand, in traditional potential energy systems, the depth of the potential well is a key
parameter governing the quantum states of particles: for a finite-depth potential well, an increase
in potential depth enhances the localization of particles, reduces the tunneling probability of the
wave function, and even converts unbound states into bound states \cite{Shankar1994}. For a near-infinitely
deep potential well, it is known that unbound states are confined to the spectral region
of $E<2t$ \cite{Liu2023}, yet how an increase in potential depth affects unbound states remains an unsolved problem.
On the other hand, with the development of non-Hermitian quantum mechanics, the combination of non-Hermitian
effects and Anderson localization has been found to induce a series of exotic quantum phenomena, such as non-Hermitian
Anderson localization \cite{Longhi2019PRL,ShuChenPRB2019}, non-Hermitian mobility edges \cite{TongLiuPRB2020,YanXiaPRB2020,XuPRB2021,PengPRB2023,XuePRL2022,JiangNJP2025}, non-Hermitian mobility edge rings \cite{Li2024}.
In the Hermitian system with a near-infinitely deep potential well, the mobility edge is known to be a real spectral line
that separates bound states from unbound states \cite{Liu2023}. However, it remains an open question whether the 
introduction of non-Hermitian effects will induce diverse spectral features and modify the fate of unbound states.
Based on these considerations, in this study, we take the quasiperiodic near-infinitely
deep potential model as the research object and systematically investigate the effects of enhanced potential
well depth and non-Hermitian effects on the unbound states of the system. In the present lattice quasiperiodic models, 
we identify unbound states with spatially extended eigenstates characterized by a vanishing IPR and a zero Lyapunov exponent, 
whereas bound states correspond to localized eigenstates with a finite IPR and a positive Lyapunov exponent. 
This identification allows us to address two central questions: whether unbound states survive when the near-infinitely 
deep potential is further deepened, and whether non-Hermitian gain and loss destroy or reshape the corresponding 
unbound-state regions. We show that unbound states persist in the deeper potential model, although their energy 
window is shifted to $-2t-V<E<2t-V$. In the non-Hermitian extensions, unbound states survive but appear in mixed 
bound-unbound phases within analytically determined real-energy intervals.

The rest of this paper is organized as follows. Section.~\ref{S2} analyzes the fate of unbound states in a 
deeper near-infinitely deep potential. Section.~\ref{S3} extends the discussion to non-Hermitian potentials 
and identifies the resulting mixed bound-unbound phases. A summary is given in Sec.~\ref{S4}.

\begin{figure}[htp]
		\centering
		\includegraphics[width=0.5\textwidth]{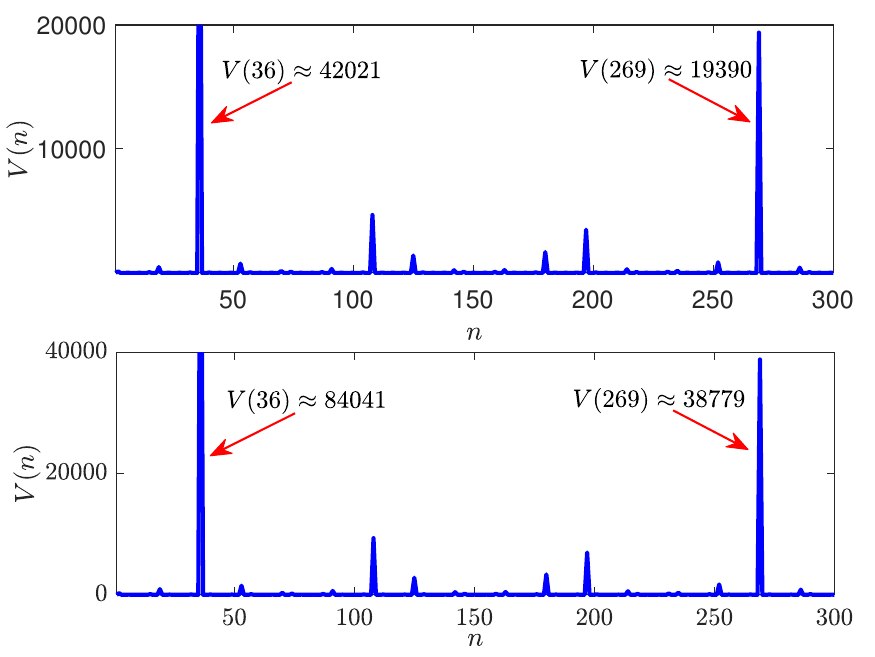}
		\caption{(Color online)
		The sketch of the near-infinitely deep potential wells
(a) $V_{1}(n)/V=\sec^{2}(2\pi\alpha n)$.
(b) $V_{2}(n)/V=\frac{1+\sin(2\pi\alpha n)^2}{\cos(2\pi\alpha n)^2}$. The comparison 
illustrates how the modified potential enhances the depth while preserving the quasiperiodic near-singular structure. }	\label{f1}
\end{figure}

\section{The deeper potential model}\label{S2}
{Liu and Xia studied a one-dimensional quasiperiodic model with the near-infinitely deep on-site 
potential $V_{1}(n)/V=\sec^{2}(2\pi\alpha n)$, where $V$ is the tunable potential strength and 
$\alpha=(\sqrt{5}-1)/2$ \cite{Liu2023}. As illustrated in Fig.~\ref{f1}(a), the potential can reach extremely 
large values at certain lattice sites, which motivates the term near-infinitely deep potential. For nearest-neighbor 
hopping, this model supports unbound states in the energy region $E<2t$, with $t$ taken as the energy unit. 
We now ask whether these unbound states survive when the potential is made even deeper. To this end, 
we introduce the modified potential $V_{2}(n)/V=\frac{1+\sin(2\pi\alpha n)^2}{\cos(2\pi\alpha n)^2}$, 
whose profile is shown in Fig.~\ref{f1}(b). The additional numerator enhances the potential depth while 
retaining the same quasiperiodic near-singular structure.

Assuming that particles in this deeper potential are allowed to hop between
nearest-neighbor lattice sites, the system can be described by the Hamiltonian
given below:
\begin{equation}
\hat{H}=\sum_{n}t\hat{c}_{n+1}^{\dagger}\hat{c}_{n}+H.c.+\sum_{n}V_2(n)\hat{c}_{n}^{\dagger}\hat{c}_{n},
\end{equation}
with $V_2(n)=V\frac{1+\sin^2(2\pi\alpha n)}{\cos^2(2\pi\alpha n)}$.

Figure~\ref{f2} presents the unbound-bound phase diagram when the system size is $L = 1597$.
The colors represent the magnitude of the inverse participation ratio (IPR). The IPR
of the $j$-th normalized wave function $|\psi^{j}\rangle=\sum_{n}\phi(n)\hat{c}_{n}^{\dag}|0\rangle$
is defined as ${\rm IPR} = \sum_{n}|\phi(n)|^4$ \cite{IPR_1}. This quantity provides a direct localization 
diagnostic: for an extended unbound state, ${\rm IPR}\sim 1/L$ and therefore vanishes in the thermodynamic limit, 
whereas for a localized bound state, the IPR remains finite \cite{Liu2023,IPR_2,IPR_3}.
As is evident from the figure, increasing the potential well depth only reduces the energy
range of the unbound states without causing their disappearance. The unbound states occur
in the energy domain $E_{c2}<E <E_{c1}$ with $E_{c1}=2t-V$ and $E_{c2}=-2t-V$,
while the bound states occur in the energy domains $E > E_{c1}$ and $E<E_{c2}$. 
To visualize the characteristics of unbound states,
Fig.~\ref{f3} shows the distributions of wave functions for four distinct energies at potential
parameter $V = 0.5t$, with energies located on either side of the critical energy $E_{c1}$.
As seen from Fig.~\ref{f3}, for $E < E_{c1}$ (Figs.~\ref{f3}(a), \ref{f3}(b)), the wave functions
are extended across the entire space, corresponding to unbound states.
In contrast, for $E > E_{c1}$ (Figs.~\ref{f3}(c), \ref{f3}(d)), the wave functions are strongly
localized in finite spatial regions and decay to zero at the boundaries, indicative of bound states.
The results are in agreement with the IPR analysis. A further finite-size scaling analysis of representative states would provide an additional confirmation that the low-IPR states obey the expected ${\rm IPR}\sim 1/L$ behavior in the thermodynamic limit.
\begin{figure}[htp]
		\centering
		\includegraphics[width=0.5\textwidth]{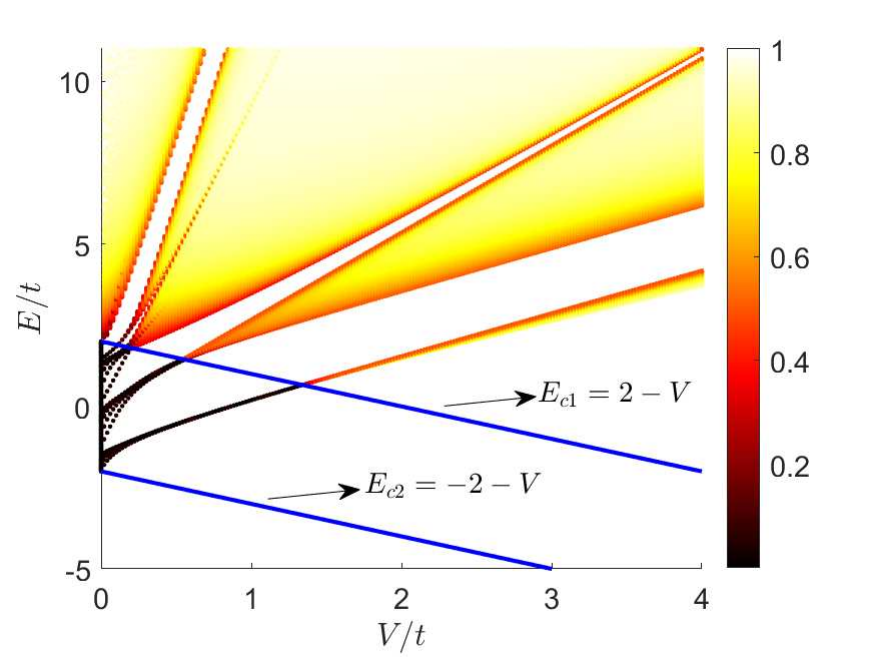}
		\caption{(Color online)
		The energies $E$ versus $V$ of the deeper potential model. The colors denote
the values of the IPR. The two blue solid lines denote the critical energies $E_{c1}=2t-V$ and
$E_{c2}=-2t-V$, respectively. Within the energy domain $\left[E_{c2},E_{c1}\right]$, there are unbound
states. Above $E_{c1}$, there are bound states. The persistence of low-IPR states inside this interval demonstrates that increasing the potential depth does not eliminate unbound states. The system size is $L=1597$.
	}	\label{f2}
\end{figure}

\begin{figure}[htp]
		\centering
		\includegraphics[width=0.5\textwidth]{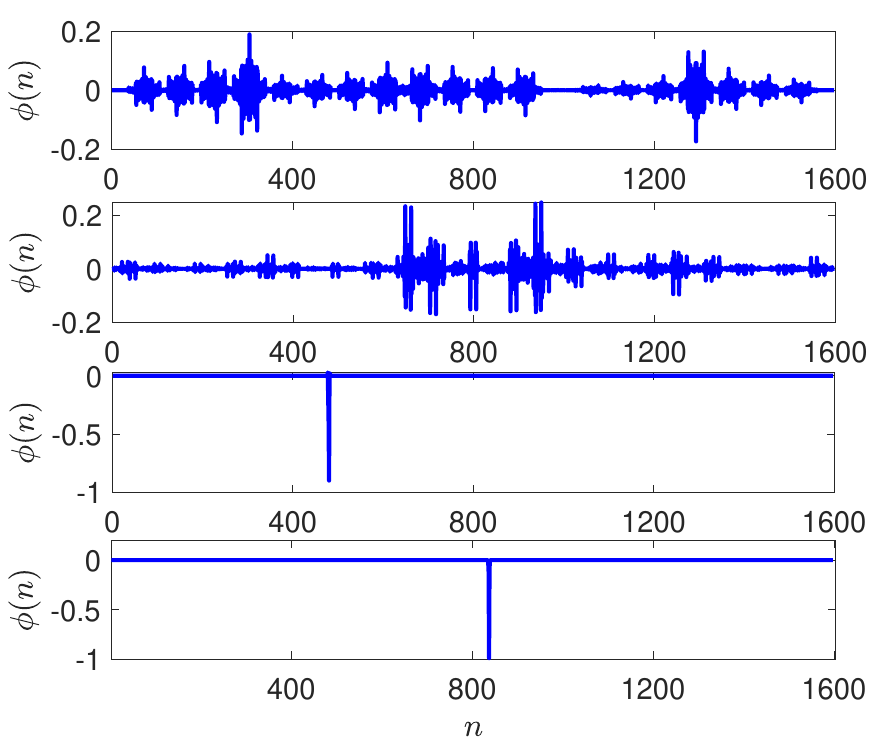}
		\caption{(Color online)
		Four wave functions of the deeper potential model.
(a) The unbound state at $E=-0.5593t$;
(b) The unbound state at $E=-0.5568t$;
(c) The bound state at $E=3.1698t$;
(d) The bound state at $E=12.8646t$.
The system size is $L=1597$.
	}	\label{f3}
\end{figure}

%By scaling the IPR, the fractal dimension $D =-\frac{\ln IPR}{\ln L}$
%can be obtained, and this can also determine whether the quantum state is a unbound state.
%For extended unbound states, it is known that IPR is proportional to $1/L$. Therefore, the
%corresponding $D$ approaches $1$. For localized bound states, the $D$ tends to $0$.
We argue that the unbound-bound transition can be exactly proved by analyzing the
Lyapunov exponent (LE) $\gamma$. For an unbound state, $\gamma=0$, while for a bound state,
$\gamma>0$. The $\gamma$ can be determined from the product
of the transfer matrices. With wave function $|\psi^{j}\rangle=\sum_{n}\phi(n)\hat{c}_{n}^{\dag}|0\rangle$ and energy $E$,
the Schr\"{o}dinger equation of amplitudes $\phi_{j}$ in transfer matrix form is written
as
\begin{equation}
\left(
\begin{array}{c}
  \phi_{n+1} \\
  \phi_{n}
\end{array}
\right)=T_{n}\left(
\begin{array}{c}
  \phi_{n} \\
  \phi_{n-1}
\end{array}
\right)
\end{equation}
with
\begin{equation}
T_{n}=\left(
\begin{array}{cc}
  \frac{E-V(n)}{t} & -1 \\
  1 & 0
\end{array}
\right).
\end{equation}

The LE of a wave function can be computed by
\begin{equation}
\gamma(E)=\lim_{L\rightarrow\infty}\frac{1}{L}\ln \parallel \prod^{L}_{n=1}T_{n}\parallel,
\end{equation}
where $\parallel \cdot \parallel$ represents the matrix norm,
which is given by the maximum absolute eigenvalue of the matrix.

It is noted that the transfer matrix $T_{n}$ can be expressed as
the product of two parts, namely
\begin{equation}
T_{n}=A_{n}B_{n}
\end{equation}
with
\begin{equation}
A_{n}=\frac{1}{\cos^2(2\pi\alpha n+\theta)}
\end{equation}
and
\begin{equation}
B_{n}=\left[
\begin{array}{cc}
  \frac{E\cos^{2}(2\pi\alpha n+\theta)-V-V\sin^2(2\pi\alpha n +\theta)}{t} & -\cos^{2}(2\pi\alpha n+\theta) \\
  \cos^2(2\pi\alpha n+\theta) & 0
\end{array}
\right]
\end{equation}

Then the $\gamma$ can be expressed as
\begin{equation}
\gamma(E)=\gamma^{A}(E)+\gamma^{B}(E),
\end{equation}
where
\begin{equation}
\begin{aligned}
\gamma^{A}(E)&=\lim_{L\rightarrow \infty}\frac{1}{L}\ln\prod^{L}_{n=1}|\frac{1}{\cos^2(2\pi\alpha n+\theta)}| \\
&=\frac{1}{2\pi} \int^{2\pi}_{0} \ln|\sec^2(\theta)|d\theta \\
&=2\ln 2.
\end{aligned}
\end{equation}

To calculate $\gamma^{B}(E)$, one should employ Avila's global theory \cite{Avila2015}.
The procedure starts with an analytical continuation of the phase, namely,
$\theta \rightarrow \theta+i\epsilon$. As $\epsilon \rightarrow \infty$,
the direct evaluation of $B_{n}$ leads to
\begin{equation}
B_{n}=\frac{1}{4}e^{-i4\pi\alpha n-i2\theta+2\epsilon}\left[
\begin{array}{cc}
  \frac{E+V}{t} & -1 \\
  1 & 0
\end{array}
\right]
\end{equation}

Then, we obtain $\gamma^{B}(E)=2\epsilon-2\ln2+\max\{\ln|\frac{\frac{E+V}{t}\pm \sqrt{\left(\frac{E+V}{t}\right)^2-4}}{2}|\}$.
According to Avila's global theory, $\gamma^{B}(E)$ is a convex and piecewise linear function with
integer slope. The quantization of the acceleration, together with this convex piecewise-linear structure, 
fixes the physical Lyapunov exponent once the large-$\epsilon$ asymptotic form is known. It can be concluded 
that the Lyapunov exponent can be determined when $\epsilon$ returns to $0$. Therefore, the LE of a wave function is
\begin{equation}
\begin{aligned}
\gamma(E)&=\gamma^{A}(E)+\gamma^{B}_{\epsilon=0}(E) \\
&=\max\{\ln|\frac{\frac{E+V}{t}\pm \sqrt{\left(\frac{E+V}{t}\right)^2-4}}{2}|\},
\end{aligned}
\end{equation}

From the above equation, we know that when $|E+V|>2t$, $\gamma(E)>0$,
while when $|E+V|<2t$, $\gamma(E)\equiv 0$. Therefore, the bound states
exist in the energy domains $E>2t-V$ and $E<-2t-V$ and the unbound states are located
in the energy domain $-2t-V<E<2t-V$. In fact, from the IPR results, we know that
below $E=2t-V$, there are already unbound states. Thus, in this deeper potential
model, $E=2t-V$ can be regarded as the zero potential energy surface.

\section{Non-Hermitian effect on unbound states}\label{S3}
The existence of unbound states in near-infinitely deep quasiperiodic potentials
 challenges the conventional expectation that an increasingly deep potential necessarily 
enhances confinement. The Hermitian analysis above shows that the fate of these states is 
governed by analytically determined Lyapunov-exponent boundaries rather than by the 
potential depth alone. We now extend this framework to non-Hermitian systems, where 
gain and loss can generate complex spectra and modify localization properties. The central 
question is whether such non-Hermitian effects destroy the unbound states or reorganize them 
into new spectral-localization structures.

We first investigate the effects of non-Hermiticity on the Liu-Xia model. With
on-site gain and dissipation effects, the non-Hermitian potential is expressed as
$V^{\rm NH}_{1}(n)=\frac{V}{\cos(2\pi\alpha n+\theta+ih)^2}\equiv V\sec(2\pi\alpha n+\theta+ih)^2$.
For the non-Hermitian models considered below, the IPR is calculated from normalized right eigenstates. 
Since the eigenenergy is generally complex, we write it as $\mathcal{E}=E+i\eta$ and use $E\equiv {\rm Re}(\mathcal{E})$ 
to denote the real-part energy unless otherwise specified. The Schr\"{o}dinger equation is then expressed in the following form:
\begin{equation}
\left(
\begin{array}{c}
  \phi_{n+1} \\
  \phi_{n}
\end{array}
\right)=T^{\rm NH1}_{n}\left(
\begin{array}{c}
  \phi_{n} \\
  \phi_{n-1}
\end{array}
\right)
\end{equation}
with the transfer matrix
\begin{equation}
T^{\rm NH1}_{n}=\left(
\begin{array}{cc}
  \frac{E-V^{\rm NH1}_{1}(n)}{t} & -1 \\
  1 & 0
\end{array}
\right).
\end{equation}

Decomposing the matrix $T^{\rm NH1}_{n}$ into a product of two
parts as $T^{\rm NH1}_{n}=A^{\rm NH1}_{n}B^{\rm NH1}_{n}$, we have
\begin{equation}
A^{\rm NH1}_{n}=\sec^{2}(2\pi\alpha n+\theta+ih)
\end{equation}
and
\begin{equation}
B^{\rm NH1}_{n}=\left[
\begin{array}{cc}
  \frac{E\cos^{2}(2\pi\alpha n+\theta+ih)-V}{t} & -\cos^{2}(2\pi\alpha n+\theta+ih) \\
  \cos^{2}(2\pi\alpha n+\theta+ih) & 0
\end{array}
\right].
\end{equation}

Then, the LE of \textcolor{red}{the} non-Hermitian system, namely $\gamma^{\rm NH1}(E)$, is expressed as
\begin{equation}
\gamma^{\rm NH1}(E)=\gamma^{A}_{\rm NH1}(E)+\gamma^{B}_{\rm NH1}(E),
\end{equation}
where
\begin{equation}
\begin{aligned}
\gamma^{A}_{\rm NH1}&=\lim_{L\rightarrow\infty}\frac{1}{L}\ln\prod^{L}_{n=1}|\sec^{2}(2\pi\alpha n+\theta+ih)| \\
&=\frac{1}{2\pi}\int^{2\pi}_{0}\ln|\sec^{2}(\theta+ih)|d\theta \\
&=2\ln2-2h.
\end{aligned}
\end{equation}

To calculate $\gamma^{B}_{\rm NH1}(E)$, we still employ Avila's
global theory. After performing the analytic continuation on $B^{\rm NH1}_{n}$,
that is, $\theta\rightarrow \theta+i\epsilon$, we have
\begin{equation}
B^{\rm NH1}_{n}(\epsilon\rightarrow\infty)=\frac{1}{4}e^{-i4\pi\alpha n-i2\theta+2h+2\epsilon}
\left[
\begin{array}{cc}
  \frac{E}{t} & -1 \\
  1 & 0
\end{array}
\right].
\end{equation}

Then, $\gamma^{B}_{\rm NH1}(E)$ is given by
\begin{equation}
\gamma^{B}_{\rm NH1}(E)=2\epsilon+2h-2\ln2 +\max\{\ln|\frac{\frac{E}{t}\pm\sqrt{\frac{E^2}{t^2}-4}}{2}|\}.
\end{equation}
As mentioned before, the LE of $B^{\rm NH1}_{n}$ can be determined when
$\epsilon$ returns to $0$. Accordingly, the Lyapunov exponent $\gamma^{\rm NH1}(E)$
is given as
\begin{equation}\label{Eq_gamma_NH1}
\gamma^{\rm NH1}(E)=\max\{\ln|\frac{\frac{E}{t}\pm\sqrt{\frac{E^2}{t^2}-4}}{2}|\}.
\end{equation}
According to Eq.~(\ref{Eq_gamma_NH1}), we know that when $|E|>2t$, $\gamma^{\rm NH1}(E)>0$, while
$|E|<2t$, $\gamma^{\rm NH1}(E)\equiv 0$. This result is highly similar to that of the Hermitian Liu-Xia model,
except that the energy $E$ here refers to the real part of the complex energy,
which will be manifested in the subsequent IPR results. The analytical results predict that the unbound states will
exist in the real-energy domain $-2t<E<2t$, while the bound states will exist in the real-energy
domains $E>2t$ and $E<-2t$. For convenience, we mark $E^{\rm NH1}_{c1}=2t$ and $E^{\rm NH1}_{c2}=-2t$.
Thus, the analytical boundaries should be compared with the real parts of the numerical eigenenergies.

\begin{figure}[htp]
		\centering
		\includegraphics[width=0.5\textwidth]{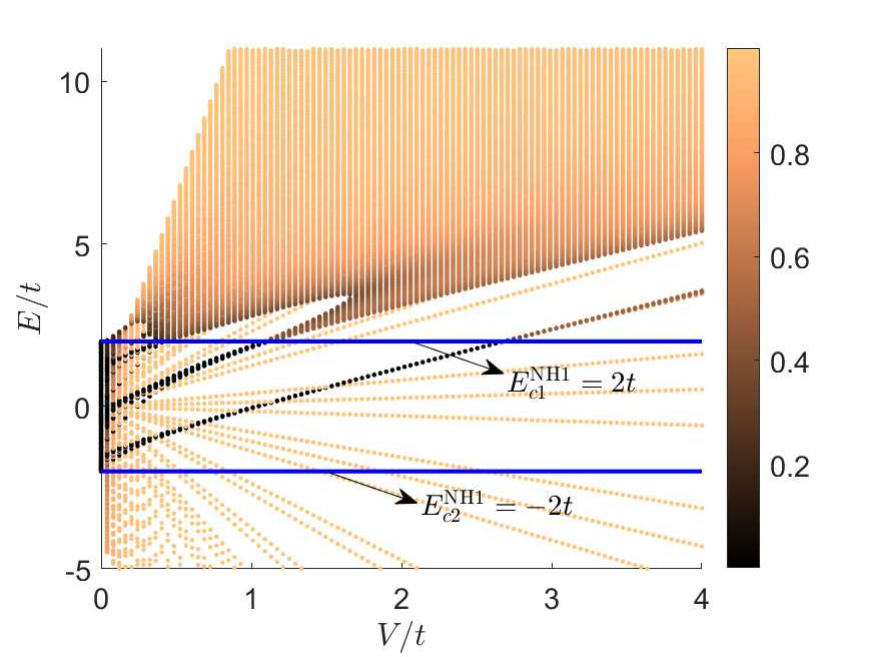}
		\caption{(Color online)
		The energies $E$ versus $V$ of the non-Hermitian Liu-Xia model. The colors denote
the values of the IPR. The two blue solid lines denote the critical energies $E^{\rm NH1}_{c1}=2t$ and
$E^{\rm NH1}_{c2}=-2t$, respectively. The energy domain $\left[E_{c2},E_{c1}\right]$ is a mixed region
where unbound states and bound states coexist. Above $E_{c1}$ and below $E_{c2}$, there are
bound states. The coexistence of low- and high-IPR states within the same real-energy interval signals the emergence of a mixed bound-unbound phase. The system size is $L=1597$.
	}	\label{f4}
\end{figure}

\begin{figure}[htp]
		\centering
		\includegraphics[width=0.5\textwidth]{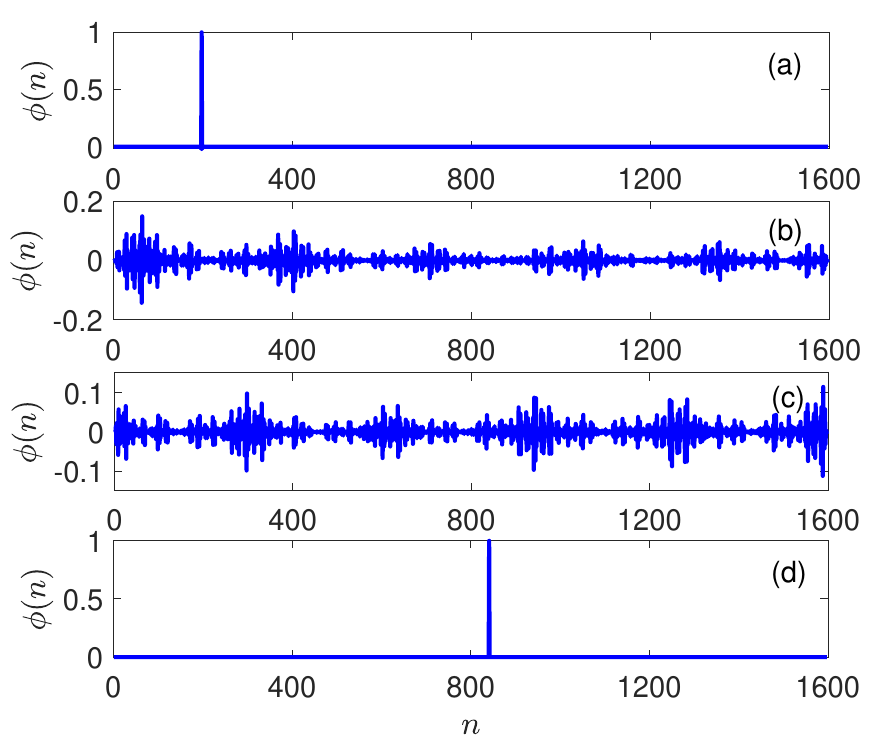}
		\caption{(Color online)
		Four wave functions of the non-Hermitian Liu-Xia model.
(a) The bound state at $E=-45.7662t$;
(b) The unbound state at $E=-0.7628t$;
(c) The unbound state at $E=-0.7625t$;
(d) The bound state at $E=1.1534t$.
The parameters used are $h=0.1$,$V=0.5t$, and $L=1597$.
	}	\label{f5}
\end{figure}

In Fig.~\ref{f4}, we present the real part of energies $E$ (arranged in ascending order) of
the non-Hermitian Liu-Xia model as a function of the potential parameter $V$. The color
coding indicates the magnitude of the IPR, and the blue lines mark the critical energies
$E^{\rm NH1}_{c1}$ and $E^{\rm NH1}_{c2}$ obtained from analytical derivations. Similar
to the Hermitian case, the energy region $E > E^{\rm NH1}_{c1}$ still accommodates bound states.
Differently, bound states emerge in the lower real-energy domain with $E < E_{c2}$. Furthermore,
it can be seen that unbound states do not vanish due to non-Hermiticity, but remain present in the energy window
$E^{\rm NH1}_{c2} < E < E^{\rm NH1}_{c1}$. Different from the Hermitian systems, the conventional
non-Hermitian mobility-edge and intermediate phase systems, in which unbound states fully occupy one or several entire
energy domains \cite{Liu2023,TongLiuPRB2020,XuPRB2021,PengPRB2023,IPR_1,IPR_2,IPR_3}, unbound states here appear in the form of mixed phase consisting
of both unbound and bound states without a regular pattern. It is readily seen that the real
part energies of unbound states are embedded in the real-part energies of bound states.
This mixed phase is distinct from a conventional mobility-edge structure: the real-energy interval 
predicted by the Lyapunov exponent provides the possible region for unbound states, but non-Hermiticity 
causes localized and extended eigenstates to coexist irregularly inside this interval.

Figures~\ref{f5}(a)-\ref{f5}(d) show the spatial distribution of the wave functions
under four different energies. In Fig.~\ref{f5}(a), the real part of the energy
corresponding to the wave function is located below the critical energy $E^{\rm NH1}_{c2}$.
It can be observed that its spatial distribution state is localized and it is a bound state.
The real-part energies of the wave functions in Figs.~\ref{f5}(b)-\ref{f5}(c) are located within
the energy domain $E^{\rm NH1}_{c2} < E < E^{\rm NH1}_{c1}$. The distributions in
Figs.~\ref{f5}(b) and \ref{f5}(d) are extended, corresponding to unbound states, while the distribution in
Fig.~\ref{f5}(c) is localized, corresponding to a bound state. This result clearly demonstrates that
in non-Hermitian cases, within the energy domain $E^{\rm NH1}_{c2} < E < E^{\rm NH1}_{c1}$,
bound states and unbound states coexist, which is consistent with the results of IPR.
This also indicates that the analytical solution derived based on Avila's global theory
can predict both the energy domain of unbound states and the energy domain of pure bound states.

Next, we study the non-Hermitian effect on the deeper potential model. Considering
on-site gain and loss in $V_{2}(n)$, the resulting non-Hermitian potential $V^{\rm NH2}_{2}(n)$ is given by
\begin{equation}
\begin{aligned}
V^{\rm NH2}_{2}(n)&=\frac{V\left[1+\sin^{2}(2\pi\alpha n+\theta+ih)\right]}{\cos^{2}(2\pi\alpha n+\theta+ih)} \\
& = V\left[\sec^{2}(2\pi\alpha n+\theta+ih)+\tan^{2}(2\pi\alpha n+\theta+ih)\right]
\end{aligned}
\end{equation}
Accordingly, the Schr\"{o}edinger equation is denoted as
\begin{equation}
\left(
\begin{array}{c}
  \phi_{n+1} \\
  \phi_{n}
\end{array}
\right)=T^{\rm NH2}_{n}\left(
\begin{array}{c}
  \phi_{n} \\
  \phi_{n-1}
\end{array}
\right)
\end{equation}
with the transfer matrix
\begin{equation}
T^{\rm NH2}_{n}=\left(
\begin{array}{cc}
  \frac{E-V^{\rm NH2}_{2}(n)}{t} & -1 \\
  1 & 0
\end{array}
\right).
\end{equation}

Following the previous strategy, here we again express
$T^{\rm NH2}_{n}$ as the product of two parts, i.e.,
$T^{\rm NH2}_{n}=A^{\rm NH2}_{n}B^{\rm NH2}_{n}$, and thus we
have
\begin{equation}
\gamma^{\rm NH2}(E)=\gamma^{A}_{NH2}(E)+\gamma^{B}_{NH2}(E).
\end{equation}

It is noted that $A^{\rm NH1}_{n}= A^{\rm NH2}_{n}$; therefore,
we have
\begin{equation}
\gamma^{A}_{\rm NH2}=\gamma^{A}_{NH1}\equiv 2\ln2-2h.
\end{equation}

To obtain $\gamma^{B}_{\rm NH2}$, we again employ Avila's global theory.
After performing the analytic continuation ($\theta\rightarrow \theta+i\epsilon$)
on the matrix $B^{\rm NH2}_{n}$, its expression in the infinite-$\epsilon$ limit reads
\begin{equation}
B^{\rm NH2}_{n}(\epsilon\rightarrow \infty)=\frac{1}{4}e^{-i4\pi\alpha n-i2\theta+2h+2\epsilon}
\left[
\begin{array}{cc}
  \frac{E+V}{t} & -1 \\
  1 & 0
\end{array}
\right].
\end{equation}
Then, $\gamma^{B}_{\rm NH2}$ is given by
\begin{equation}
\gamma^{B}_{\rm NH2}=2\epsilon+2h-2\ln2+\max\{\ln|\frac{\frac{E+V}{t}\pm\sqrt{(\frac{E+V}{t})^2-4}}{2}|\}
\end{equation}

\begin{figure}[htp]
		\centering
		\includegraphics[width=0.5\textwidth]{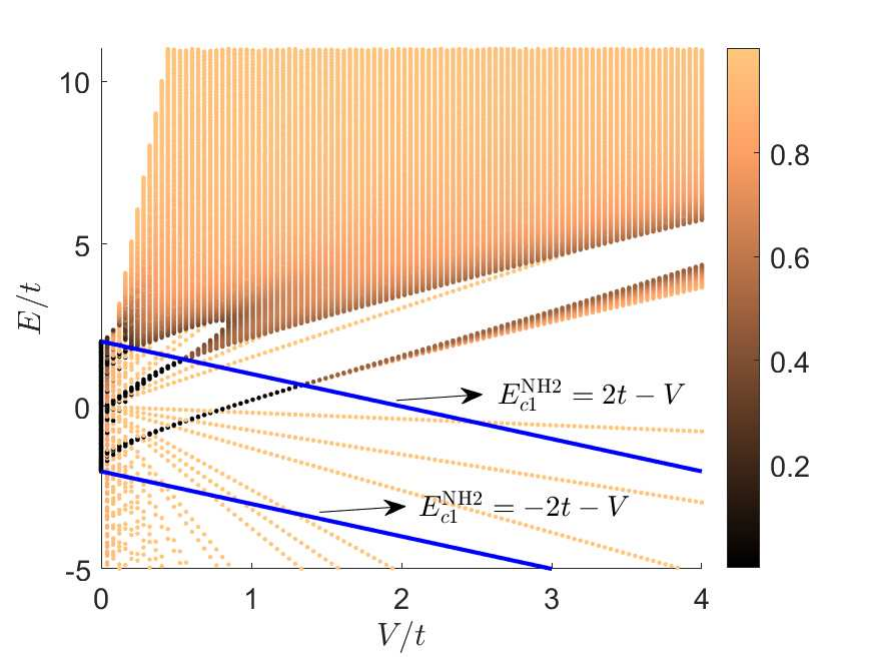}
		\caption{(Color online)
		The real-part energies $E$ versus $V$ of the non-Hermitian deeper potential model.
The colors denote the values of the IPR. The two blue solid lines denote the critical
energies $E^{\rm NH2}_{c1}=2t-V$ and
$E^{\rm NH2}_{c2}=-2t-V$, respectively. The energy domain $\left[E_{c2},E_{c1}\right]$ is a mixed region
where unbound states and bound states coexist. Above $E_{c1}$ and below $E_{c2}$, there are
bound states. The mixed interval follows the shifted analytical boundaries inherited from the deeper potential. The system size is $L=1597$ and $h=0.1$.
	}	\label{f6}
\end{figure}

\begin{figure}[htp]
		\centering
		\includegraphics[width=0.5\textwidth]{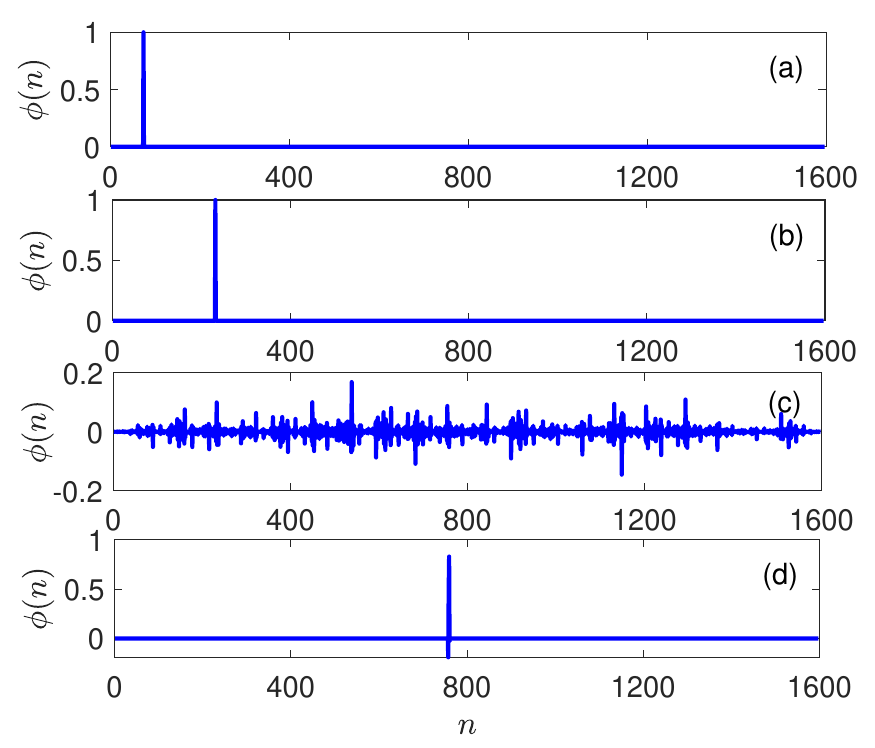}
		\caption{(Color online)
		Four wave functions of the non-Hermitian deeper potential model.
(a) The bound state at $E=-3.1524t$;
(b) The bound state at $E=-0.7393t$;
(c) The unbound state at $E=0.1940t$;
(d) The bound state at $E=2.2617t$.
The parameters used are $h=0.1$, $V=t$, and $L=1597$.
	}	\label{f7}
\end{figure}

Since $\gamma^{B}_{\rm NH2}$ is a convex function that is linear in sections and has an integer slope,
\textcolor{red}{it} can be concluded that the LE can be obtained when $\epsilon$ goes back to $0$.
Therefore, the LE $\gamma^{\rm NH2}(E)$ is
\begin{equation}\label{Eq_gamma_NH2}
\gamma^{\rm NH2}(E)=\max\{\ln|\frac{\frac{E+V}{t}\pm\sqrt{(\frac{E+V}{t})^2-4}}{2}|\}
\end{equation}
From the above Eq.~(\ref{Eq_gamma_NH2}), we know that when $|E+V|>2t$, $\gamma^{\rm NH2}(E)>0$;
while when $|E+V|<2t$, $\gamma^{\rm NH2}(E)=0$. The analytical results predict that the unbound
states exist in the real-energy domain $-2t-V<E<2t-V$, while the bound states appear in
the real-energy domains $E>2t-V$ \textcolor{red}{and} $E<-2t-V$. This expression is consistent
with the result of the Hermitian theory. However, it should be emphasized that for non-Hermitian models,
the energy here refers to its real part. This is fully reflected in the subsequent IPR calculations.
For convenience, we mark $E^{\rm NH2}_{c1}=2t-V$ and $E^{\rm NH2}_{c2}=-2t-V$.

In Fig.~\ref{f6}, we plot the real-part energies $E$ (arranged in ascending order) as
a function of the potential parameter $V$. The color indicates the magnitude of the IPR, and
the blue lines mark the critical energies $E^{\rm NH2}_{c1}$ and $E^{\rm NH2}_{c2}$ obtained
from analytical derivations. Similar to the Hermitian case, bound states still occupy
the energy domain $E > E^{\rm NH2}_{c1}$. Differently, bound states emerge in the lower real
part energy domain with $E < E^{\rm NH2}_{c2}$. In this regard, the distribution characteristics
of the bound states are quite similar to that of the non-Hermitian Liu-Xia model. Furthermore, it is evident that the
unbound states do not vanish due to non-Hermiticity, but remain present in the real-part energy window
$E^{\rm NH2}_{c2} < E < E^{\rm NH2}_{c1}$.  Consistent with the non-Hermitian Liu-Xia model,
the unbound states here still appear in the form of a mixed phase consisting of both unbound
and bound states without a regular pattern. It is readily seen that the real-part energies of unbound states
are embedded in the real-part energies of bound states as well. The reappearance of the mixed phase 
indicates that the non-Hermitian reconstruction of localization properties is robust against the deepening 
of the quasiperiodic potential.

The unbound states, bound states, and mixed phase can all be reflected by the spatial
distributions of the wave functions. Figures~\ref{f7}(a)-\ref{f7}(d) show the spatial
distributions of the wave functions under four different real-part energies. In Fig.~\ref{f7}(a),
the real-part energy corresponding to the wave function lies below the critical
energy $E^{\rm NH2}_{c2}$. It can be observed that its spatial distribution is localized,
indicating a bound state. The wave functions in Figs.~\ref{f7}(b)-\ref{f7}(c) fall within
the real-part energy domain $E^{\rm NH2}_{c2} < E < E^{\rm NH2}_{c1}$. Concretely, the
distribution in Fig.~\ref{f7}(b) is localized, corresponding to a bound state,
while that in Fig.~\ref{f7}(c) is extended, corresponding to an unbound state.
For fixed system parameters, the coexistence of both bound and unbound states
in the energy spectrum characterizes the mixed phase. When the real-part energy
is larger than $E^{\rm NH2}_{c1}$, Fig.~\ref{f7}(d) shows that the wave function
distribution is localized, corresponding to a bound state. These results clearly
demonstrate that in the non-Hermitian deeper potential model, bound states and
unbound states coexist within the energy domain $E^{\rm NH2}_{c2} < E < E^{\rm NH2}_{c1}$,
which is consistent with the IPR results. This also indicates that the analytical
solution derived based on Avila's global theory can predict both the real-part energy
domain of unbound states and that of pure bound states.

\section{Summary}\label{S4}
In summary, we have investigated the fate of unbound states in one-dimensional quasiperiodic models with near-infinitely deep potentials. For the deeper Hermitian model, numerical IPR calculations and wave-function profiles show that increasing the potential depth does not remove the unbound states. Instead, Lyapunov-exponent analysis based on Avila's global theory gives the shifted unbound-state window $-2t-V<E<2t-V$, in agreement with the numerical phase diagram. Thus, the existence of unbound states is not determined simply by the absolute depth of the potential, but by the analytical energy window fixed by the Lyapunov exponent.

We have further extended both the Liu-Xia model and its deeper counterpart to non-Hermitian quasiperiodic potentials with gain and loss. In these cases, the relevant analytical boundaries are defined in terms of the real part of the complex eigenenergy. The Lyapunov exponent identifies the real-energy intervals in which unbound states can appear, while numerical IPR results reveal that these intervals become mixed bound-unbound phases rather than purely unbound regions. In contrast to conventional mobility-edge scenarios, localized and extended eigenstates coexist irregularly within the same real-energy window. Our results therefore show that non-Hermiticity does not destroy the unbound states in near-infinitely deep potentials, but reshapes their spectral organization and produces mixed localization structures. These findings provide a theoretical basis for further studies of bound-unbound transitions in extreme quasiperiodic and non-Hermitian systems.

This research is supported by Zhejiang Provincial Natural Science Foundation of China under Grant No. LQN25A040012,
the start-up fund from Xingzhi College, Zhejiang Normal University, and the National Natural Science Foundation of China under Grant No. 12174346.

\bibliography{reference}

\end{document}